\documentclass[conference]{IEEEtran}
\IEEEoverridecommandlockouts

\usepackage{cite}
\usepackage{amsmath,amssymb,amsfonts}
\usepackage{graphicx}
\usepackage{subfloat}
\usepackage{subcaption}
\usepackage{multirow} 
\usepackage{makecell}
\usepackage{textcomp}
\usepackage{algorithm}
\usepackage{algorithmicx}
\usepackage{algpseudocode} 
\usepackage[T1]{fontenc}
\usepackage{xcolor}
\usepackage{amssymb}
\usepackage{booktabs}
\usepackage{color}
\usepackage{mathrsfs}
\usepackage{bm}
\usepackage{hyperref}

\usepackage{soul}
\def\BibTeX{{\rm B\kern-.05em{\sc i\kern-.025em b}\kern-.08em
    T\kern-.1667em\lower.7ex\hbox{E}\kern-.125emX}}
\begin{document}

\title{User Localization via Active Sensing with Electromagnetically Reconfigurable Antennas\\
}

\author{Ruizhi Zhang, Yuchen Zhang,~\textit{IEEE Member}, Ying Zhang,~\textit{IEEE Member}
	\thanks{Ruizhi Zhang and Ying Zhang are with University of Electronic Science and Technology of China, China (202122011309@std.uestc.edu.cn, zhying@uestc.edu.cn). Yuchen Zhang is with the Electrical and Computer Engineering Program, Computer, Electrical and Mathematical Sciences and Engineering (CEMSE), King Abdullah University of Science and Technology (KAUST), Thuwal 23955-6900, Kingdom of Saudi Arabia (yuchen.zhang@kaust.edu.sa).
	}
}

\maketitle

\begin{abstract}
	This paper presents an end-to-end deep learning framework for electromagnetically reconfigurable antenna (ERA)-aided user localization with active sensing, where ERAs provide additional electromagnetic reconfigurability to diversify the received measurements and enhance localization informativeness. 
	To balance sensing flexibility and overhead, we adopt a two-timescale design: the digital combiner is updated at each stage, while the ERA patterns are reconfigured at each substage via a spherical-harmonic representation. The proposed mechanism integrates attention-based feature extraction and LSTM-based temporal learning, enabling the system to learn an optimized sensing strategy and progressively refine the UE position estimate from sequential observations. Simulation results show that the proposed approach consistently outperforms conventional digital beamforming-only and single-stage sensing baselines in terms of localization accuracy. These results highlight the effectiveness of ERA-enabled active sensing for user localization in future wireless systems.
\end{abstract}


\section{Introduction}

Accurate and low-latency user localization is an essential capability for future wireless systems, supporting applications such as intelligent transportation, industrial automation, and digital twins \cite{Trevlakis_OJCOMS_6G}. 
Especially, rich physical information in the angular, delay, and frequency domains provide additional localization potential in wideband systems\cite{Rappaport_Access_mmWave, Xue_COMST_BeamMgmt}. 
Nevertheless, practical deployments often operate in complex environments where multiple propagation paths overlap in time and angle, resulting in ambiguity and increased sensitivity to noise, which makes robust localization challenging \cite{Trevlakis_OJCOMS_6G, Zafari_COMST_IndoorSurvey, Wymeersch_VTM_RISLocMap}.

Recently, electromagnetically reconfigurable antennas (ERAs) have emerged as a promising hardware paradigm for enhancing localization by introducing controllable electromagnetic degrees of freedom at the transceiver \cite{FAS_PartI,FAS_PartII,Fadakar_HybridCodebook_ERFAS}. 
By electronically adjusting antenna-domain properties such as radiation pattern, ERAs can generate multiple radiation modes without mechanical movement, enabling better manipulation of the array response \cite{Zheng_ERA6G,Ojaroudi_Survey2020}. 
In particular, the reconfigurability of pattern diversifies the received signals beyond conventional fixed-pattern arrays, enriches the delay-angle observation space, and thereby improving the informativeness of observations for localization \cite{Fadakar_HybridCodebook_ERFAS,Hakkarainen_CROWNCOM2014}. 
Moreover, jointly optimizing ERAs pattern with a digital beamformer can further enhance sensing performance and localization quality by coherently exploiting electromagnetic- and digital domain flexibility \cite{Fadakar_HybridCodebook_ERFAS,Zheng_arXiv_TriHybridMUPrecoding}.

In parallel, active sensing has attracted increasing interest as a powerful approach to sequentially adapt sensing configurations based on multi-stage observations. Compared with wide scanning, it progressively focuses sensing resources on promising configurations, which can lower computation and measurement cost while achieving accurate localization \cite{Zhang_TWC_ActiveRIS,LiYu_CL_MultipathActiveRIS}.
By leveraging intermediate feedback, active sensing enables a progressive reduction of localization error and supports an exploration-exploitation behavior, where early stages perform wide-area probing and later stages concentrate sensing resources on the most informative regions \cite{Zhang_TWC_ActiveRIS,Wymeersch_VTM_RISLocMap}. 
Such adaptation is particularly beneficial in multipath environments, since it can mitigate ambiguity caused by overlapping paths \cite{LiYu_CL_MultipathActiveRIS,Alexandropoulos_LWC_MultiRIS}. 
Learning-based designs are especially appealing for active sensing, since the optimal sequential strategy typically involves high-dimensional variables and functional policies, making it a challenging optimization problem \cite{SohrabiYu_JSAC_DeepActiveBeam,Zhang_TWC_ActiveRIS}.

Motivated by these developments, this paper studies RA-aided user localization via deep learning-based active sensing in a wideband system. 
To balance sensing flexibility and configuration overhead, we adopt a two-timescale protocol where the digital combiner is updated at the stage level while the RA patterns are reconfigured at each substage. We develop an end-to-end framework that integrates an attention-based feature extractor for compact wideband observation embedding and an LSTM cell for temporal representation learning across stages. The learned hidden state is then used to predict the next-stage sensing configuration and refine the UE position estimate. 
Simulation results demonstrate the effectiveness of learning-driven RA active sensing for robust localization.

The remainder of this paper is organized as follows. Section~II presents the system model and problem formulation. Section~III introduces the proposed deep learning-based active sensing mechanism. Section~IV reports simulation results, followed by conclusions in Section~V.

\section{System Model and Problem Formulation}

\subsection{System Model}
We consider a wireless uplink localization task in an environment consisting of a single-antenna UE and an AP equipped with $N$ RAs. The AP is capable of jointly configuring the receive digital beamformer and the radiation pattern of the RA to enhance the localization accuracy. 
The UE transmits OFDM pilots over $M$ active subcarriers, and the received signal propagates through a multipath channel, which is modeled as the superposition of $P$ distinct propagation paths. To balance the adaptation flexibility and signaling overhead, we adopt a two-timescale protocol: the digital beamformer is updated at the stage level, whereas the RA radiation pattern can be adjusted for each substage to fine-probe the channel and acquire informative measurements. 
Accordingly, the entire localization process can be divided into $T$ stages. 
For the $t$-th stage, the AP applies a stage-wise digital beamformer $\mathbf{w}_t\in\mathbb{C}^{N}$, expressed as $\mathbf{w}_t=\left[w_t(1),w_t(2),\ldots,w_t(N)\right]^T$, which satisfies the norm constraint $\|\mathbf{w}_t\|_2^2\leq P_{\rm max}$, where $P_{\rm max}$ denotes the maximum power budget. Meanwhile, the RA radiation pattern is configured on each substage to provide additional sensing diversity, and can be modeled via a truncated real spherical-harmonic expansion, as follows.

In each stage $t\in\{1,\ldots,T\}$, the UE transmits $L$ consecutive pilot symbols.  For the $n$-th AP antenna, we represent the amplitude gain toward the direction
$\bm u=(\sin\theta\cos\phi,\sin\theta\sin\phi,\cos\theta)$ as~\cite{Fadakar_HybridCodebook_ERFAS,Zheng_arXiv_TriHybridMUPrecoding,Chen_arXiv_TriHybridISAC_ERA}
\begin{equation}
	G^n_{t,l}(\theta,\phi)
	\;\triangleq\;
	\boldsymbol{\gamma}(\theta,\phi)^{\top}\bm c^n_{t,l},
	\label{eq:pattern-def}
\end{equation}
where $\boldsymbol{\gamma}(\theta,\phi)\in\mathbb{R}^{K}$ stacks the first
$K=U^2+2U+1$ spherical-harmonic basis functions up to degree $U$, and
$\bm c^n_{t,l}\in\mathbb{R}^{K}$ denotes the corresponding coefficient vector
during the $l$-th pilot of stage $t$. To place all radiation patterns on a common energy scale, we enforce the normalization $\int_{0}^{\pi}\!\!\int_{-\pi}^{\pi}
\big(G^n_{t,l}(\theta,\phi)\big)^2 \sin\theta\,\mathrm d\phi\,\mathrm d\theta
\;=\; 1$,
which is equivalent to $\|\bm c^n_{t,l}\|_2^2 = 1$.
We define the block-diagonal coefficient matrix that collects the coefficients of all antennas for the $l$-th substage in stage $t$ as~\cite{Zheng_ERA6G}
\begin{equation}
	\mathbf F^{\rm cof}_{t,l}
	\;\triangleq\;
	\operatorname{blkdiag}\!\big(\bm c^1_{t,l},\bm c^2_{t,l},\ldots,\bm c^N_{t,l}\big)
	\in\mathbb{R}^{(NK)\times N}.
	\label{eq:Fcof}
\end{equation}
For any arrival direction $(\theta,\phi)$, we further define $\boldsymbol{\Gamma}^{\rm AP}(\theta,\phi) \;\triangleq\; \mathbf I_{N}\otimes \boldsymbol{\gamma}(\theta,\phi)^{\top} \in\mathbb{R}^{N\times(NK)}$,
then the per-antenna RA gains toward $(\theta,\phi)$ can be written as
\begin{equation}
	\begin{split}
		\mathbf D_{t,l}(\theta,\phi)
		&\;\triangleq\;
		\boldsymbol{\Gamma}^{\rm AP}(\theta,\phi)\,\mathbf F^{\rm cof}_{t,l} \\
		&\;=\;
		\operatorname{diag}\!\big(G^1_{t,l}(\theta,\phi),\ldots,G^N_{t,l}(\theta,\phi)\big).
	\end{split}
	\label{eq:D-theta-phi}
\end{equation}

Based on the geometric channel model, each path $p$ can be characterized by a complex gain $\alpha_p$, a propagation delay $\tau_p$, and an arrival direction
$\bm u_p=(\sin\theta_p\cos\phi_p,\sin\theta_p\sin\phi_p,\cos\theta_p)$.
With the AP steering vector $\bm a(\bm u_p)\in\mathbb{C}^N$ given by
$[\bm a(\bm u_p)]_n=\exp\!\big(j\frac{2\pi}{\lambda}\bm r_n^\top\bm u_p\big)$,
the uplink frequency-domain channel vector on subcarrier $m\in\{1,\ldots,M\}$ during the $l$-th pilot of stage $t$ is modeled as
\begin{equation}
	\mathbf h_{t,l}^m
	\;=\;
	\sum\nolimits_{p}
	\alpha_p\,
	\mathbf D_{t,l}(\theta_p,\phi_p)\,\bm a(\bm u_p)\,
	e^{-\mathrm j2\pi \tau_p (m-1)\triangle f }
	\;\in\mathbb C^{N},
	\label{eq:channel-vector-ofdm}
\end{equation}
where $\triangle f$ denotes the subcarrier spacing.

Finally, we consider an OFDM pilot symbol transmitted over $M$ subcarriers. Let $\mathbf{s}_l \in \mathbb{C}^M$ denote the $l$-th transmitted pilot vector over $M$ subcarriers at each stage.
After digital beamforming, the received frequency-domain observation vector is given by
\begin{equation}
	\mathbf{y}_{t,l}
	=
	\big[
	\mathbf{w}_{t}^{H}\mathbf{h}_{t,l}^{1}[\mathbf{s}_l]_1,\ \ldots,\ 
	\mathbf{w}_{t}^{H}\mathbf{h}_{t,l}^{M}[\mathbf{s}_l]_M
	\big]^{T}
	+\mathbf{n}_{t,l},
	\label{eq:obs-ofdm}
\end{equation}
where $\mathbf{n}_{t,l} \in \mathbb{C}^M$ is additive noise with $\mathbf{n}_{t,l} \sim \mathcal{CN}(\mathbf{0},\sigma^2\mathbf{I}_M)$.

\subsection{Problem Formulation}\label{sec:problem}

Building on the above active sensing system model, our goal is to estimate the UE position by exploiting the received pilots across $T$ stages.
We define the stage-level observation matrix as $\mathbf Y_t\triangleq\big[\mathbf y_{t,1},\mathbf y_{t,2},\ldots,\mathbf y_{t,L}\big]\in\mathbb C^{M\times L}$.
At the beginning of stage $t$, the AP configures the stage-wise digital beamformer $\mathbf w_t$ and the substage-wise RA radiation patterns via harmonic coefficients $\{\bm c^n_{t,l}\}_{n=1,l=1}^{N,L}$, based on all observations collected from the previous stages, i.e., $\{\mathbf Y_{\tau}\}_{\tau=1}^{t-1}$. This induces a functional stage-wise configuration policy:
\begin{equation}
	\Big\{\mathbf w_{t},\ \{\bm c^n_{t,l}\}_{n=1,l=1}^{N,L}\Big\}
	=
	\mathcal{G}^{t}\!\left(\left\{\mathbf Y_{\tau}\right\}_{\tau=1}^{t-1}\right),
	\label{eq:policy-G-wideband}
\end{equation}
where $\mathcal{G}^{t}(\cdot)$ determines the sensing configuration of stage $t$.

After collecting all observations $\{\mathbf Y_t\}_{t=1}^{T}$, the AP produces the final UE location estimate $\widehat{\bm p}$ through a localization mapping $\mathcal{F}_{\rm loc}(\cdot)$:
\begin{equation}
	\widehat{\bm p}
	=
	\mathcal{F}_{\rm loc}\!\left(\left\{\mathbf Y_t\right\}_{t=1}^{T}\right).
	\label{eq:policy-F-wideband}
\end{equation}

The RA-aided wideband uplink localization problem is to design the configuration policies $\{\mathcal{G}^{t}(\cdot)\}_{t=1}^{T}$ together with the estimator $\mathcal{F}_{\rm loc}(\cdot)$ to minimize the mean-squared localization error (MSE), subject to the digital beamformer and RA pattern constraints. Specifically, we formulate
\begin{subequations}\label{prob:p1_wideband}
	\begin{align}
		(\mathcal{P}1):\quad
		\min_{\{\mathcal{G}^{t}\}_{t=1}^{T},\,\mathcal{F}}
		\quad & \mathbb{E}\!\left[\left\|\widehat{\bm p}-\bm p\right\|_2^2\right]
		\label{prob:p1_obj_wideband}\\
		\text{s.t.}\quad
		& \|\mathbf w_{t}\|_2^2 \le P_{\rm max},\quad \forall t,
		\label{prob:p1_power_wideband}\\
		& \|\bm c^n_{t,l}\|_2^2 = 1,\quad \forall t,\,\forall l,\,\forall n,
		\label{prob:p1_pattern_wideband}
	\end{align}
\end{subequations}
However, problem~\eqref{prob:p1_wideband} is highly non-convex and involves optimizing over function spaces, making a globally optimal analytical solution intractable. Therefore, in the following section, we develop a data-driven solution based on a deep learning architecture that learns the stage-wise adaptation $\{\mathcal{G}^{t}\}$ and the final estimator $\mathcal{F}_{\rm loc}$ from training data.

\section{Proposed Mechanism}\label{sec:proposed}
\begin{figure}[!t]
	\centering 
	\vspace*{-0.5em}
	\includegraphics[width=3.39in]{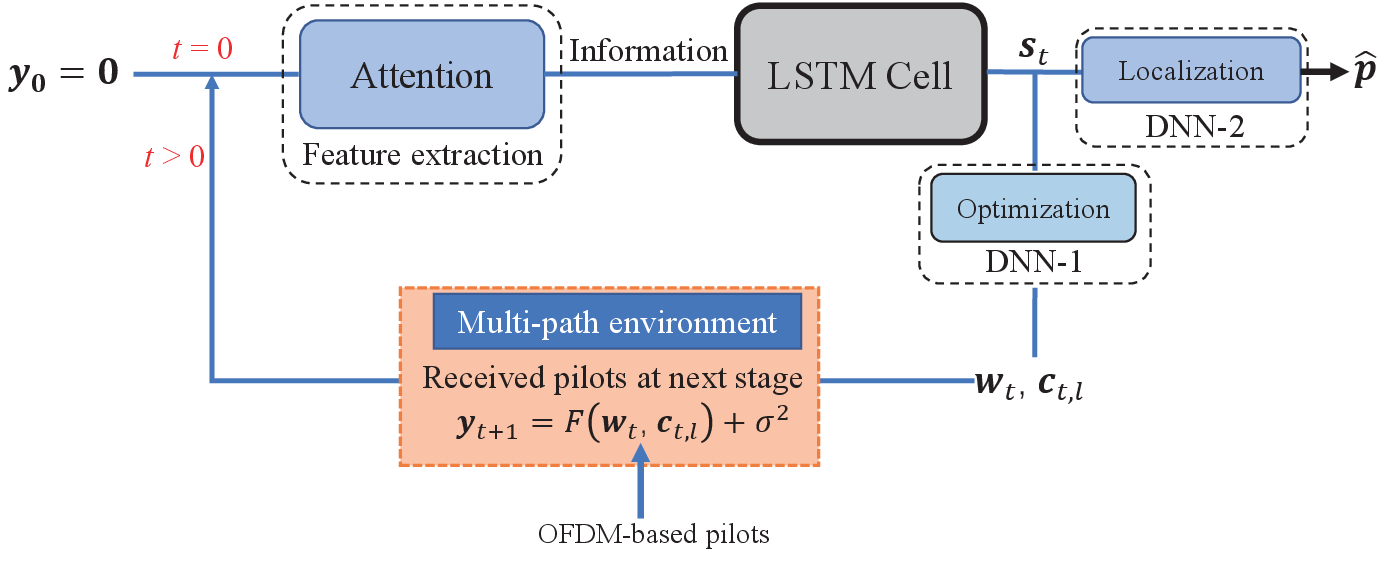} 
	\vspace*{-0.5em}
	\caption{The PRA-aided uplink active sensing system.}
	\vspace*{-1.5em}
	\label{fig:framework}
\end{figure}
Our proposed method employs an LSTM-based closed-loop framework to estimate the UE location by exploiting the received pilots over multiple subcarriers. 
To manage the complexity arising from processing high-dimensional multi-subcarrier observation, we group the $L$ consecutive pilots within each stage into a stage-level input. As shown in Fig.~\ref{fig:framework}, the proposed architecture comprises four main components: (i) an attention-based feature extractor, (ii) an LSTM cell for temporal representation learning, (iii) an optimization network (DNN-1) for active sensing configuration, and (iv) a localization network (DNN-2) for position estimation.

In stage $t$, the AP collects $L$ pilot observations $\{\mathbf y_{t,l}\}_{l=1}^{L}$, where each $\mathbf y_{t,l}\in\mathbb C^{M}$ corresponds to the $M$ subcarrier frequency-domain observation after digital beamforming. We stack these vectors into a stage-level measurement matrix	$\mathbf Y_t \;\triangleq\; 	\big[\mathbf y_{t,1},\mathbf y_{t,2},\ldots,\mathbf y_{t,L}\big] \in\mathbb C^{M\times L}$, which preserves both the frequency diversity and probing diversity.

However, directly feeding $\mathbf Y_t$ into an LSTM may be prone to overfitting due to the large input dimension. Therefore, we employ an attention-based feature extractor to learn compact embeddings from $\mathbf Y_t$. Specifically, we first convert complex-valued measurements into real-valued features via real-imaginary concatenation, i.e., $\widetilde{\mathbf Y}_t \;\triangleq\; \big[\Re(\mathbf Y_t);\Im(\mathbf Y_t)\big]	\in\mathbb R^{2M\times L}$, and then apply an attention module to adaptively aggregate the $L$ pilot-wise observations
\begin{equation}
	\mathbf z_t
	\;=\;
	\mathcal A(\widetilde{\mathbf Y}_t)
	\in\mathbb R^{d},
	\label{eq:attention_embedding}
\end{equation}
where $\mathcal A(\cdot)$ denotes the attention-based feature extractor and $d<2ML$ is the embedding dimension. 

Then the extracted embedding $\mathbf z_t$ is fed into an LSTM cell to capture the temporal dependence across stages.
Let $\mathbf s_t$ denote the LSTM hidden state. The recurrent update is expressed as
\begin{equation}
	\mathbf s_t
	\;=\;
	\mathcal L(\mathbf z_t,\mathbf s_{t-1}),
	\label{eq:lstm_update}
\end{equation}
where $\mathcal L(\cdot)$ denotes the LSTM transition, and $\mathbf s_t$ is a compact memory summarizing all past information, enabling AP to progressively refine both configuration and position estimation~\cite{LiYu_CL_MultipathActiveRIS}.
To realize active sensing, we feed the LSTM state $\mathbf s_t$ into an optimization network $\mathcal O(\cdot)$, which outputs the AP configuration for the next stage
\begin{equation}
	\Big\{\mathbf w_{t+1},\ \{\bm c^n_{t+1,l}\}_{n=1,l=1}^{N,L}\Big\}
	=
	\mathcal O(\mathbf s_t),
	\label{eq:opt_net}
\end{equation}
The predicted configuration is then deployed at the AP receiver to acquire the next-stage observations $\mathbf Y_{t+1}$ based on~\eqref{eq:obs-ofdm}, forming a closed-loop stage-wise sensing mechanism.

In parallel, the LSTM state $\mathbf s_t$ is also connected to a localization network, which maps the learned temporal representation to a UE position estimate
\begin{equation}
	\widehat{\bm p}_t
	=
	\mathcal F_{\rm loc}(\mathbf s_t),
	\label{eq:loc_net}
\end{equation}
After $T$ stages, the final estimate can be taken as $\widehat{\bm p}\triangleq\widehat{\bm p}_{T}$~\cite{LiYu_CL_MultipathActiveRIS,SohrabiYu_JSAC_DeepActiveBeam}.
The overall architecture is trained end-to-end using supervised learning with a stage-wise weighted MSE. Specifically, since the early stages are primarily used for exploration, while the later stages focus on refined localization, we assign smaller weights to the earlier stages and larger weights to later stages. The training objective is given by
\begin{equation}
	\min \ \mathbb E\!\left[\sum_{t=1}^{T}\beta_t\,
	\big\|\widehat{\bm p}_{t}-\bm p\big\|_2^2\right],
	\label{eq:weighted_mse_loss}
\end{equation}
where $\{\beta_t\}_{t=1}^{T}$ are nonnegative weights which increases with $t$ to emphasize the localization accuracy at later stages.

\section{Experiment and Results}
\begin{figure}[!t]
	\centering
	\vspace*{-0.5em}
	\begin{subfigure}[t]{0.240\textwidth}
		\centering
		\includegraphics[width=\linewidth]{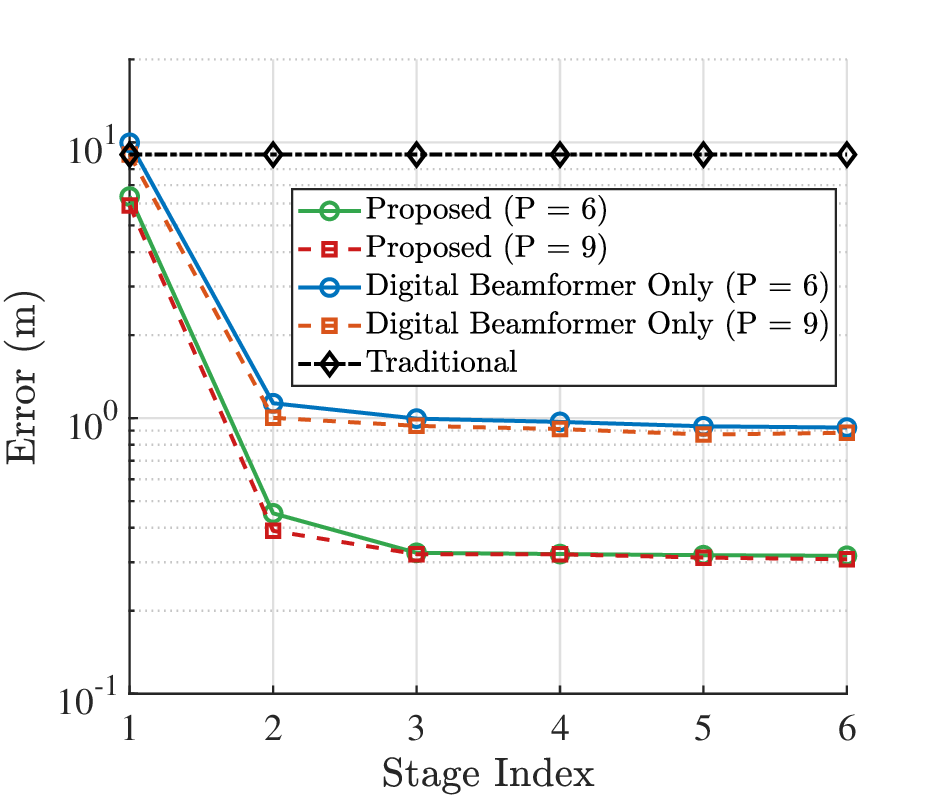}
		\caption{}
		\label{fig:com_stage}
	\end{subfigure}
	\begin{subfigure}[t]{0.240\textwidth}
		\centering
		\includegraphics[width=\linewidth]{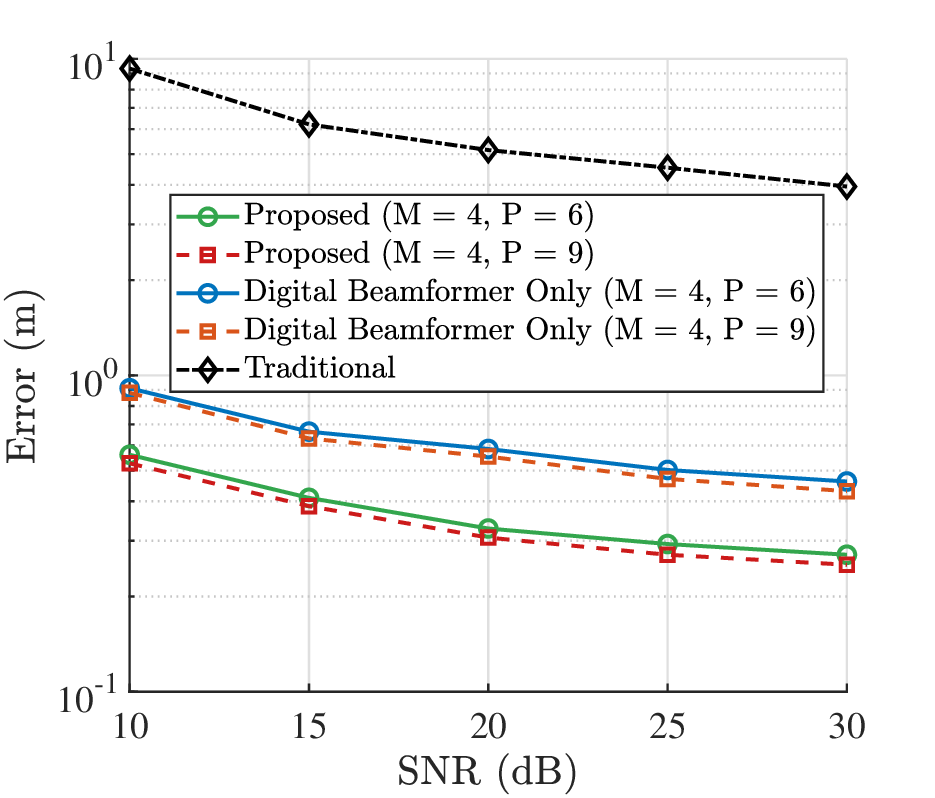}
		\caption{}
		\label{fig:com_snr}
	\end{subfigure}
\vspace*{-0.5em}
	\caption{Localization performance comparison: (a) localization RMSE versus stage variation. (b) localization RMSE versus SNR.}
	\label{fig:com_stage_snr}
\end{figure}

\begin{figure}[!t]
	\vspace*{-1em}
	\centering 
	\includegraphics[width=3.1in]{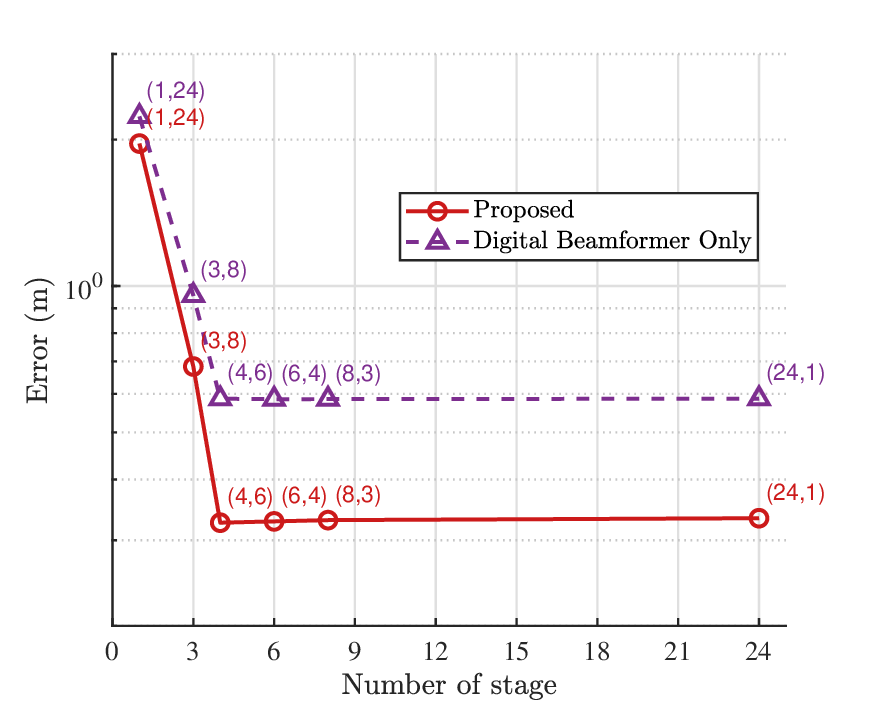} 
	\caption{Localization RMSE versus stage and pilot allocation under fixed total pilots.}
	\vspace*{-1.0em}
	\label{fig:com_total}
\end{figure}

\begin{figure*}[!t]
	\centering
	\begin{subfigure}[t]{0.823\textwidth}
		\centering
		\includegraphics[width=\linewidth]{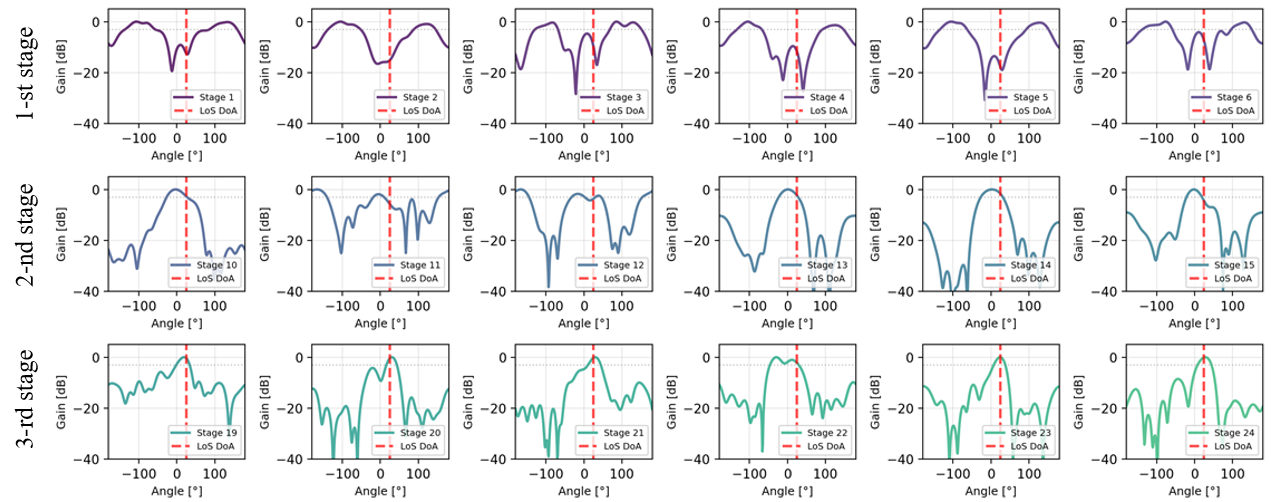}
		\caption{}
		\label{fig:beam_proposed}
	\end{subfigure}
	\begin{subfigure}[t]{0.169\textwidth}
		\centering
		\includegraphics[width=\linewidth]{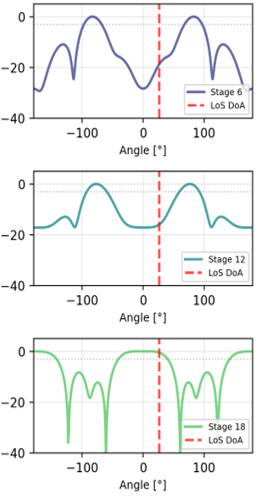}
		\caption{}
		\label{fig:beam_digital}
	\end{subfigure}
	\caption{Stage-wise synthesized beampatterns under \(3\) stages with \(6\) pilots per stage: (a) Proposed ERA-assisted active sensing. (b) Digital beamformer only.}
	\vspace*{-1.5em}
	\label{fig:beam_compare}
\end{figure*}
In this section, we evaluate the proposed method described in~\ref{sec:proposed} for the active sensing localization system. We consider a wideband OFDM pilot transmission occupying $M=256$ subcarriers. The carrier frequency is set to $f_c=30$~GHz and the subcarrier spacing is $\Delta f=240$~kHz~\cite{Wei_PRS_Sensing_JSC}. The AP is equipped with $N=25$ ERAs arranged as a uniform planar array with half-wavelength spacing. We adopt a per-substage per-pilot setting, where one pilot is transmitted and one observation is collected in each substage. The number of resolvable propagation paths is set to $P=3$. 
For the network, the stage-wise feature extractor is achieved by a lightweight transformer encoder. Specifically, the encoder first projects $\widetilde{\mathbf Y}_t$ into a latent space with model dimension $d_{\rm model}=128$, and then applies a multi-head self-attention layer with $H=4$ heads, followed by a pooling network producing an $256$ dimension embedding vector. The temporal dependency across stages is captured by a LSTM cell with hidden state dimension $256$, whose hidden state is shared by the optimization head and the localization head~\cite{Vaswani_NIPS_Transformer,Hochreiter_NC_LSTM}.

We generate a dataset by sampling UE locations $\bm p\sim\mathcal{U}([-30,30]\times[-30,30])$ and drawing the corresponding channel parameters (e.g., path gains, delays, and angles) according to the adopted multipath channel model. In total, we generate $30000$ samples, among which $90\%$ are used for training and the remaining $10\%$ are used for validation.

To demonstrate the advantages of the proposed active sensing framework and the ERA-enabled pattern reconfiguration, we compare our method with two representative baselines at two levels: (i) active sensing versus one-shot sensing, and (ii) within active sensing, ERA-assisted versus digital beamforming only. The first baseline considers \emph{digital beamforming only} under the active sensing setup, where the AP employs a conventional antenna array to perform multi-stage active sensing. In this case, only the stage-wise digital beamformer $\mathbf w_t$ is optimized. The second baseline corresponds to \emph{traditional one-shot sensing}, where the AP performs sensing in a single stage and optimizes only digital beamformer, without active sensing.

Fig.~\ref{fig:com_stage_snr} illustrates how the localization performance varies with different parameters. As shown in Fig.~\ref{fig:com_stage}, the localization error decreases rapidly with the stage variation. In the initial stage, the sensing configuration focus on wide area exploration, leading to a relatively large error. After receiving early pilots, the proposed method quickly reduces the localization uncertainty, followed by a gradual refinement and convergence in later stages. In contrast, the digital beamforming only baseline converges to a higher error floor, while the traditional single-stage sensing remains unchanged due to the lack of sequential adaptation.  Fig.~\ref{fig:com_snr} further illustrates the final localization error versus SNR. The proposed method consistently outperforms both baselines across all SNR values, with more pronounced gains in the low-to-moderate SNR regime. This confirms that actively optimizing sensing configuration based on intermediate observations improves both noise robustness and measurement efficiency.

Fig.~\ref{fig:com_total} evaluates different designs under the same total pilot budget, by varying the number of stages and the number of pilots transmitted per stage. The proposed method benefits significantly from splitting the sensing process into multiple stages, since intermediate feedback enables progressive refinement of the UE location. Compared with the single-stage setting, allocating the same total pilots into a moderate number of stages yields much lower localization error, while further increasing the stage number provides diminishing returns due to the reduced per-stage measurement overhead. Overall, the proposed scheme consistently outperforms the digital beamforming only baseline for all allocation strategy, demonstrating superior pilot efficiency under constrained sensing resources.

Fig.~\ref{fig:beam_compare} visualizes the synthesized beampatterns across different stages, under the setting of 3 stages with 6 pilots per stage. As shown in Fig.~\ref{fig:beam_proposed}, in the early stage, the pattern of the ERA-based active sensing is relatively broad, which facilitates wide-angle exploration under limited prior information. As the active sensing progresses and more pilot observations are accumulated, the beampattern gradually becomes more concentrated around the LoS direction, showing a clear convert from exploration to exploitation. In contrast, the digital beamforming only baseline lacks sufficient adaptivity without ERA reconfiguration as shown in Fig.~\ref{fig:beam_digital}, leading to reduced angular selectivity and lower sensing efficiency. This confirms that the proposed ERA-based active sensing mechanism can effectively utilize received signal to allocate sensing energy toward the most informative angular region.

\section{Conclusion}

This paper proposed a deep-learning-based active sensing framework for ERA-aided user localization in a wideband system, leveraging substage-wise radiation-pattern reconfiguration to enhance measurement diversity. 
A two-timescale protocol and an end-to-end architecture combining attention-based embedding with LSTM-based temporal learning were developed to enable active sensing and progressive position refinement. 
Simulation results verified that the proposed method consistently outperforms conventional beamforming-only baseline under various SNR conditions and pilot budgets. 
Future work will extend the proposed framework to multi-user localization with pilot scheduling and interference-aware sensing, and to mobile-user tracking where temporal dynamics can be exploited for continuous position refinement.


\begin{thebibliography}{99}
	
	\bibitem{Trevlakis_OJCOMS_6G}
	S.~E.~Trevlakis \emph{et al.},
	``Localization as a Key Enabler of 6G Wireless Systems: A Comprehensive Survey and an Outlook,''
	\emph{IEEE Open Journal of the Communications Society},
	vol.~4, pp.~2733--2801, 2023.
	
	\bibitem{Rappaport_Access_mmWave}
	T.~S.~Rappaport, S.~Sun, R.~Mayzus, H.~Zhao, Y.~Azar, K.~Wang, G.~N.~Wong,
	J.~K.~Schulz, M.~Samimi, and F.~Gutierrez,
	``Millimeter Wave Mobile Communications for 5G Cellular: It Will Work!,''
	\emph{IEEE Access}, vol.~1, pp.~335--349, 2013.
	
	\bibitem{Xue_COMST_BeamMgmt}
	Q.~Xue, C.~Ji, S.~Ma, J.~Guo, Y.~Xu, Q.~Chen, and W.~Zhang,
	``A Survey of Beam Management for mmWave and THz Communications Towards 6G,''
	\emph{IEEE Communications Surveys \& Tutorials}, vol.~26, no.~3, pp.~1520--1559, 2024.
	
	\bibitem{Zafari_COMST_IndoorSurvey}
	F.~Zafari, A.~Gkelias, and K.~K.~Leung,
	``A Survey of Indoor Localization Systems and Technologies,''
	\emph{IEEE Communications Surveys \& Tutorials}, vol.~21, no.~3, pp.~2568--2599, 2019.
	
	\bibitem{Wymeersch_VTM_RISLocMap}
	H.~Wymeersch, J.~He, B.~Denis, A.~Clemente and M.~Juntti,
	``Radio Localization and Mapping with Reconfigurable Intelligent Surfaces: Challenges, Opportunities, and Research Directions,''
	\emph{IEEE Vehicular Technology Magazine}, vol.~15, no.~4, pp.~52--61, Dec.~2020.
	
	\bibitem{FAS_PartI}
	K.-K.~Wong, W.~K.~New, X.~Hao, K.-F.~Tong, and C.-B.~Chae,
	``Fluid Antenna System---Part I: Preliminaries,''
	\emph{IEEE Communications Letters}, vol.~27, no.~8, pp.~1919--1923, Aug.~2023.
	
	\bibitem{FAS_PartII}
	K.-K.~Wong, K.-F.~Tong, and C.-B.~Chae,
	``Fluid Antenna System---Part II: Research Opportunities,''
	\emph{IEEE Communications Letters}, vol.~27, no.~8, pp.~1924--1928, Aug.~2023.
	
	\bibitem{Fadakar_HybridCodebook_ERFAS}
	A.~Fadakar, Y.~Zhang, H.~Chen, M.~F.~Keskin, H.~Wymeersch, and A.~F.~Molisch,
	``Hybrid Codebook Design for Localization Using Electromagnetically Reconfigurable Fluid Antenna System,''
	\emph{arXiv preprint,} arXiv:2508.21351, 2025.
	

	\bibitem{Zheng_ERA6G}
	P.~Zheng, R.~Wang, Y.~Zhang, M.~J.~Hossain, A.~Chaaban, A.~Shamim, and T.~Y.~Al-Naffouri,
	``Electromagnetically Reconfigurable Antennas for 6G: Enabling Technologies, Prototype Studies, and Research Outlook,''
	\emph{arXiv preprint,} arXiv:2506.00657, 2025.
	
	\bibitem{Ojaroudi_Survey2020}
	N.~Ojaroudi~Parchin, H.~J.~Basherlou, Y.~I.~A.~Al-Yasir, A. M. Abdulkhaleq, and R.~A.~Abd-Alhameed,
	``Reconfigurable Antennas: Switching Techniques---A Survey,''
	\emph{Electronics}, vol.~9, no.~2, p.~336, 2020.
	
	\bibitem{Hakkarainen_CROWNCOM2014}
	A.~Hakkarainen, J.~Werner, N.~Gulati, D.~Patron, D.~Pfeil, H.~Paaso, A.~M\"ammel\"a, K.~Dandekar, and M.~Valkama,
	``Reconfigurable Antenna Based DoA Estimation and Localization in Cognitive Radios: Low Complexity Algorithms and Practical Measurements,''
	in \emph{Proc. 9th Int. Conf. Cognitive Radio Oriented Wireless Networks and Communications (CROWNCOM)}, 2014.
	
	\bibitem{Zheng_arXiv_TriHybridMUPrecoding}
	P.~Zheng, Y.~Zhang, T.~Y.~Al-Naffouri, M.~J.~Hossain, and A.~Chaaban,
	``Tri-Hybrid Multi-User Precoding Using Pattern-Reconfigurable Antennas: Fundamental Models and Practical Algorithms,''
	\emph{arXiv preprint},
	arXiv:2505.08938, May~2025.
	
	\bibitem{Zhang_TWC_ActiveRIS}
	Z.~Zhang, T.~Jiang, and W.~Yu,
	``Localization With Reconfigurable Intelligent Surface: An Active Sensing Approach,''
	\emph{IEEE Transactions on Wireless Communications},
	vol.~23, no.~7, pp.~7698--7711, Jul.~2024.
	
	\bibitem{LiYu_CL_MultipathActiveRIS}
	Y.~Li and W.~Yu,
	``Localization in Multipath Environments via Active Sensing With Reconfigurable Intelligent Surfaces,''
	\emph{IEEE Communications Letters}, vol.~28, no.~9, pp.~2061--2065, Sep.~2024.
	
	\bibitem{Alexandropoulos_LWC_MultiRIS}
	G.~C.~Alexandropoulos, I.~Vinieratou, and H.~Wymeersch,
	``Localization via Multiple Reconfigurable Intelligent Surfaces Equipped With Single Receive RF Chains,''
	\emph{IEEE Wireless Communications Letters}, vol.~11, no.~5, pp.~1072--1076, May~2022.
	
	\bibitem{SohrabiYu_JSAC_DeepActiveBeam}
	F.~Sohrabi, Z.~Chen, and W.~Yu,
	``Deep Active Learning Approach to Adaptive Beamforming for mmWave Initial Alignment,''
	\emph{IEEE Journal on Selected Areas in Communications},
	vol.~39, no.~8, pp.~2347--2360, Aug.~2021.
	
	
	\bibitem{Chen_arXiv_TriHybridISAC_ERA}
	J.~Chen, X.~Lei, Y.~Zhang, K.~Meng, and C.~Masouros,
	``Integrated Sensing and Communication with Tri-Hybrid Beamforming Across Electromagnetically Reconfigurable Antennas,''
	\emph{arXiv preprint},
	arXiv:2510.14530, Oct.~2025.
	
	
	\bibitem{Vaswani_NIPS_Transformer}
	A.~Vaswani, N.~Shazeer, N.~Parmar, J.~Uszkoreit, L.~Jones, A.~N.~Gomez, 
	{\L}.~Kaiser, and I.~Polosukhin,
	``Attention Is All You Need,''
	in \emph{Advances in Neural Information Processing Systems (NeurIPS)}, 2017.
	
	\bibitem{Hochreiter_NC_LSTM}
	S.~Hochreiter and J.~Schmidhuber,
	``Long Short-Term Memory,''
	\emph{Neural Computation}, vol.~9, no.~8, pp.~1735--1780, 1997.
	
	\bibitem{Wei_PRS_Sensing_JSC}
	Z.~Wei, Y.~Wang, L.~Ma, S.~Yang, Z.~Feng, C.~Pan, Q.~Zhang, Y.~Wang, H.~Wu, and P.~Zhang,
	``5G PRS-Based Sensing: A Sensing Reference Signal Approach for Joint Sensing and Communication System,''
	\emph{IEEE Transactions on Vehicular Technology}, vol.~72, no.~3, pp.~3250--3263, Mar.~2023.
	
	
	
\end{thebibliography}
\end{document}